\documentstyle{elsart} 
 
\begin{document} 
 
\begin{frontmatter} 
 
\title{Shell-Model Effective Operators for Muon Capture in $^{20}$Ne} 
 
\author[jyvaeskylae]{T.\ Siiskonen}, 
\author[jyvaeskylae]{J.\ Suhonen} 
\and 
\author[oslo]{M.\ Hjorth-Jensen} 
 
\address[jyvaeskylae] 
{Department of Physics, University of Jyv\"askyl\"a, FIN-40351 Jyv\"askyl\"a, 
        Finland} 
 
\address[oslo]{Department of Physics, University of Oslo, N-0316 Oslo, Norway} 
 
\maketitle 
 
\begin{abstract} 
It has been proposed that 
the discrepancy between the partially-conserved axial-current prediction 
and the nuclear shell-model calculations of the ratio $C_{\rm P}/C_{\rm A}$ 
in the muon-capture reactions can be solved in the case of $^{28}$Si by 
introducing effective transition operators. Recently there has been 
experimental interest in measuring the needed angular correlations also in 
$^{20}$Ne. Inspired by this, we have performed a shell-model analysis employing 
effective transition operators in the shell-model formalism 
for the transition $^{20}{\rm Ne}(0^+_{\rm g.s.})+\mu^- 
\to {}^{20}{\rm F}(1^+;\, 1.057\, {\rm MeV})+\nu_\mu$. Comparison of the 
calculated capture rates with existing data supports the 
use of effective transition operators. Based on our calculations, as soon as 
the experimental 
anisotropy data becomes available, the limits for the ratio $C_{\rm P}/ 
C_{\rm A}$ can be extracted. 
\end{abstract}


\end{frontmatter}

The large energy release in the ordinary (non-radiative) capture of stopped 
negative muons by atomic nuclei probes the hadronic current much deeper than 
ordinary beta decay or electron capture. In particular, the role 
of the induced pseudoscalar coupling $C_{\rm P}$ becomes important in muon 
capture.  
Based on this, there have been many attempts in the past to extract the ratio 
of the induced pseudoscalar 
and axial-vector coupling constants, $C_{\rm P}/C_{\rm A}$, from measured 
capture rates (see e.g.\ \cite{mor60,cie76,gmi90,kuz94,joh96,sii98}) as well as 
from angular correlations of the gamma emission 
following the capture reaction ${}^A_Z{\rm X}^{}_N+\mu^-\to{}^{\phantom{AA}A} 
_{Z-1}{\rm X}_{N+1}^{'*}+\nu_\mu$ of polarized muons (see e.g.\ 
\cite{mof97,bru95,sii99}). 
 
The 
angular correlation data, available for muon capture in $^{28}$Si, has been 
in a key role in pointing out discrepancies in the shell-model calculations 
of $C_{\rm P}/C_{\rm A}$. In various shell-model  calculations (see e.g.\ \cite{mof97,sii99} 
and references therein) anomalously small values of this ratio  
($C_{\rm P}/C_{\rm A}\sim 0$) have been obtained.  
In \cite{sii99} we have proposed a method 
which, 
at least partly, lifts this discrepancy. This method is based on the use of 
effective transition operators in the shell-model formalism. Unfortunately, the 
anisotropy data is available only for $^{28}$Si and thus further testing of the 
effective-operator method has to be done in the context of measured capture 
rates or future experiments on angular correlations in the capture of polarized 
muons. At this point it is worth pointing out that the matrix elements of muon 
capture 
can also be applied to various other problems, one of the most interesting 
being the search of the scalar coupling of the hadronic current \cite{ego88}. 
 
A new measurement of the correlation coefficients of $\gamma$-radiation 
following the capture reaction in $^{28}$Si has been reported recently \cite{bru99}, 
confirming the earlier results of Refs.\ \cite{mof97} and \cite{bru95}. 
So far $^{28}$Si has 
been the only nucleus where this angular correlation data exists. In these 
experiments the parameter 
        \begin{equation} 
        \label{x} 
        x\equiv M_1(2)/M_1(-1) 
        \end{equation} 
can be extracted via the scaling coefficient $\alpha$ of the angular correlation 
between the emitted $\gamma$-radiation and the muon neutrino. This scaling 
coefficient is related to $x$ as \cite{bru95} 
        \begin{equation} 
        \label{alpha} 
        \alpha\equiv\frac{\sqrt2 x-x^2/2}{1+x^2}. 
        \end{equation} 
The quantities $M_1(-1)$ and $M_1(2)$ are linear combinations of reduced nuclear 
matrix elements and are given by 
        \begin{equation}\label{M1} 
        M_1(-1)=\sqrt{2\over3}\left\{\left({1\over3}G_{\rm P}-G_{\rm A} 
        \right)[101]+G_{\rm P}{\sqrt2\over3}[121]{C_{\rm A} 
        \over M}[011p]+{C_{\rm V}\over M}\sqrt{2\over3}[111p]\right\}, 
        \end{equation} 
        \begin{equation}\label{M2} 
        M_1(2)=\sqrt{2\over3}\left\{\left(G_{\rm A}-{2\over3}G_{\rm P}\right) 
        [121]-G_{\rm P}{\sqrt2\over3}[101]+ 
        {C_{\rm A}\over M}\sqrt2 
        [011p]+{C_{\rm V}\over M}\sqrt{2\over3}[111p]\right\}, 
        \end{equation} 
where $M$ is the nucleon mass. The definitions of the reduced nuclear matrix 
elements $[\dots ]$ can be found e.g.\ from \cite{sii98}. 
The constants 
$G_{\rm P}$ and $G_{\rm A}$ are related to the weak-interaction coupling 
constants as 
        \begin{eqnarray} 
        G_{\rm P}&=&(C_{\rm P}-C_{\rm A}-C_{\rm V}-C_{\rm M}){E_\nu\over2M},\\ 
        G_{\rm A}&=&C_{\rm A}-(C_{\rm V}+C_{\rm M}){E_\nu\over2M}. 
        \end{eqnarray} 
 
Using the expressions of Eqs.\ (\ref{M1}) and (\ref{M2}), combined with Eq.\ 
(\ref{x}), the value 
of $C_{\rm P}/C_{\rm A}$ can be extracted if the experimental value of $x$ is 
known. However, calculations with different nuclear models give very different 
predictions for this ratio. In Ref.\ \cite{bru95} the values 
$C_{\rm P}/C_{\rm A}=3.4\pm1.0$ and $C_{\rm P}/C_{\rm A}=2.0\pm1.6$ were 
extracted using the matrix elements of Refs.\ \cite{cie76} and \cite{par81}, 
respectively. In addition, the measurement of Ref.\ \cite{mof97} gives the 
estimates $C_{\rm P}/C_{\rm A}=5.3\pm2.0$ using the matrix elements of 
\cite{cie76} and $C_{\rm P}/C_{\rm A} 
=4.2\pm2.5$ using the matrix elements of \cite{par81}. 
 
The more realistic matrix elements, obtained 
from the full 1s0d shell calculation utilizing Wildenthal's USD 
interaction \cite{wil84}, yield the value of $C_{\rm P}/C_{\rm A}=0.0\pm3.2$ 
\cite{mof97,jun96}, far from the value $C_{\rm P}/C_{\rm A}\approx7$ given by 
the nuclear-model independent Goldberger-Treiman relation (see e.g.\ 
\cite{com83}) 
obtained by using the partially-conserved axial-current hypothesis (PCAC). 
The estimate given by the shell-model matrix elements is very surprising, 
since the 
USD interaction is fitted to a selected set of the 1s0d-shell spectroscopic 
data, reproducing various spectroscopic quantities like 
energy spectra, Gamow-Teller decay properties and electromagnetic properties 
(see e.g.\ \cite{car86,bro87}) very well. 
 
This anomaly, present in the shell-model calculations of $^{28}{\rm Si}(0^+ 
_{\rm g.s.})+\mu^-\to{}^{28}{\rm Al}(1^+_3)+\nu_\mu$, 
can be, at least partly, avoided by using renormalized one-body 
transition operators in the context of the shell model. In the work of 
\cite{sii99} the USD and effective interactions based on the  
recent CD-Bonn \cite{mac96} and  
Nijmegen \cite{nim94} nucleon-nucleon (NN) interaction models, yielded the interval 
$0.4\le C_{\rm P}/C_{\rm A}\le 2.7$, whereas with the renormalized transition 
operators the interval $3.4\le C_{\rm P}/C_{\rm A}\le 5.4$ was obtained, closer 
to the PCAC-prediction of this ratio.  
This value agrees also with the recent analysis of Brudanin et al.\ \cite{bru99}. 
Moreover, of special interest are the recent plans for the angular-correlation 
measurements following the capture reaction ${}_{10}^{20}{\rm Ne}(0^+_{\rm 
g.s.})+\mu^- \to{}_{\phantom{0}9}^{20}{\rm F}(1^+_1)+\nu_\mu$, as  announced 
in \cite{bru99}. In 
the present Letter we investigate this particular reaction in the shell-model 
framework with and without effective operators, and give predictions for 
the ratio $C_{\rm P}/C_{\rm A}$ using 
different sets of two-body interactions. The needed muon-capture formalism is 
treated in great detail in Ref.\ \cite{mor60} and reviewed in the shell-model 
context e.g.\ in Ref.\ \cite{sii98}.  
 
In the present shell-model calculation 
we have employed three different two-body interactions. 
In addition to the abovementioned USD interaction \cite{wil84}, we have  
derived microscopic effective interactions and operators based on 
the recent CD-Bonn meson-exchange NN interaction model  of Machleidt {\em et al.}\ 
\cite{mac96} and the Nijm-I NN interaction model of the Nijmegen group \cite{nim94}. 
These are the same interactions which were  
 employed by us in Ref.\  \cite{sii99}. 
In order to obtain effective interactions, see Ref.\ \cite{hko95} for more 
details, 
and operators for the muon capture studies, 
we use $^{16}$O as a closed-shell nucleus and define the 1s0d shell as 
the shell-model space for which the effective interactions and operators are 
derived. Based on a $G$-matrix derived for  $^{16}$O, we include all diagrams 
through third-order in $G$ and sum folded diagrams to infinite order 
employing the so-called  
$\hat{Q}$-box approach described in e.g.\ Ref.\ \cite{hko95}, in order 
to derive an effective two-body interaction for the 1s0d shell. 
In the discussions below, we will refer to these effective two-body 
interactions simply as CD-Bonn and Nijm-I interactions. 
 
The effective single-particle operators are calculated along the same lines  
as the effective interactions. In   
nuclear transitions, the quantity of 
interest is the transition matrix element between an initial state 
$\left|\Psi_i\right\rangle$ and a final state $\left|\Psi_f\right\rangle$ 
of an operator ${\cal O}$ defined as 
\begin{equation} 
               {\cal O}_{fi}= 
               \frac{\left\langle\Psi_f\right| 
               {\cal O}\left|\Psi_i\right\rangle } 
               {\sqrt{\left\langle\Psi_f | \Psi_f \right\rangle 
               \left\langle \Psi_i | \Psi_i \right\rangle}}. 
               \label{eq:effop1} 
\end{equation} 
Since we perform our calculation in a reduced space, the exact 
wave functions $\Psi_{f,i}$ are not known, only their 
projections $\Phi_{f,i}$ onto the model space. We are then confronted with the 
problem of how to evaluate ${\cal O}_{fi}$ when only the model 
space wave functions are known. In treating this problem, it is usual 
to introduce an effective operator 
${\cal O}_{fi}^{\mathrm{eff}}$, defined by 
requiring 
\begin{equation} 
           {\cal O}_{fi}=\left\langle\Phi_f\right |{\cal O}_{\mathrm{eff}} 
           \left|\Phi_i\right\rangle. 
\end{equation} 
Observe that ${\cal O}_{\mathrm{eff}}$ 
is different from the original operator ${\cal O}$. The standard 
scheme is then to employ a  
perturbative expansion for the effective operator, see e.g.\ Refs.\  
\cite{towner87,eo77}. 
 
To obtain effective one-body transition operators for muon capture, we 
evaluate all effective operator diagrams through second-order in the 
$G$-matrix obtained with  the CD-Bonn and Nijm-I interactions. Such diagrams 
are discussed in the reviews by Towner \cite{towner87} 
and Ellis and Osnes \cite{eo77}.   
Terms arising from meson-exchange currents have 
been neglected, similarly, also the possibility 
of having isobars $\Delta$ as intermediate states are omitted 
since the focus here is  on nucleonic degrees 
of freedom only. Moreover, the nucleon-nucleon potentials 
we are employing do already contain such intermediate states. 
Including $\Delta$ degrees of freedom may thus lead to a possible 
double-counting. 
Intermediate-state excitations in each diagram 
up to $6-8\hbar\omega$ in oscillator energy were included 
in order to achieve a converged result. This is also in line 
with studies of effective interactions with weak tensor 
forces \cite{sommerman}, 
such as the CD-Bonn potential employed here. 
 
The energy spectrum of $^{20}$F, emerging from our full 1s0d-shell calculation 
using $^{16}$O 
as closed-shell core, is shown in Fig.\ \ref{fig:spec}.  
The agreement with 
experiment is good.  
In particular, both the CD-Bonn and Nijm-I results are very close to 
the USD ones, and the energy of the $1^+_1$ final state of the capture reaction is 
reproduced almost exactly. The description of the spectrum of the double-even 
$^{20}$Ne nucleus by shell-model is more trivial than the description of the 
spectrum of the double-odd $^{20}$F. For this reason the agreement between the 
calculated and measured \cite{CD} energy spectra of $^{20}$Ne is excellent 
for all interactions and thus we refrain from a detailed comparison of the 
$^{20}$Ne spectra. The shell-model calculations were performed using 
the code OXBASH \cite{oxb88}. The reader should note that  
since the USD interaction is an effective 
interaction operating in the 1s0d shell only, it is not possible to calculate 
with this interaction 
the corresponding effective operators which connect to states outside the 
1s0d model space. Therefore, we have employed the effective operators obtained 
with the CD-Bonn interaction  
for the USD calculation as well. Employing those from the  
Nijm-I interaction gives similar results. 
 
The renormalization effects on the one-body transition matrix elements are 
of the order of $10-30\%$, and in almost all cases we get reduction in the 
absolute value. 
In particular, the Gamow--Teller-type single-particle matrix elements, 
corresponding to the matrix element $[101]$, reduce roughly by 10\%. However, it 
should be noted, that the radial dependence in the $[101]$ matrix element 
differs from the radial dependence of the pure Gamow--Teller matrix element. 
The resulting nuclear 
matrix elements for the transition $^{20}{\rm Ne}(0^+_{\rm g.s.})+\mu^- 
\to {}^{20}{\rm F}(1^+;\, 1.057\, {\rm MeV})+\nu_\mu$ 
are shown in Table \ref{nme}, obtained by combining the one-body transition 
matrix elements with the corresponding one-body transition densities of the 
shell-model calculation. 
\begin{table} 
\caption{The values of the reduced nuclear matrix elements (RNME). The recoil 
        matrix elements $[\dots p]$ are given in units of fm$^{-1}$.} 
\begin{tabular}{lrrrrrr}\hline 
             &\multicolumn{2}{c}{USD} &\multicolumn{2}{c}{CD-Bonn}& 
             \multicolumn{2}{c}{Nijm-I}\\\hline 
        RNME & bare & renorm & bare & renorm& bare & renorm \\ 
        \hline 
        $[101]$  &  0.0203 &  0.0218 &  0.0244 &  0.0251 &0.0249 &0.0256 \\ 
        $[121]$  &  0.0045 &  0.0028 &  0.0039 &  0.0024 &0.0043 &0.0028 \\ 
        $[101-]$ &  0.0192 &  0.0209 &  0.0233 &  0.0241 &0.0237 &0.0246\\ 
        $[121+]$ &  0.0055 &  0.0034 &  0.0048 &  0.0029 &0.0052 &0.0034\\ 
        $[111p]$ &  0.0303 &  0.0231 &  0.0286 &  0.0219 &0.0281 &0.0220\\ 
        $[011p]$ & -0.0136 & -0.0091 & -0.0165 & -0.0110 &-0.0163 &-0.0109\\ 
        \hline 
\end{tabular} 
\label{nme} 
\end{table} 
The corresponding capture rates  
obtained using the formalism of Ref.\ \cite{mor60} are shown 
in Fig.\ \ref{rates} with the experimental value of Ref.\ \cite{fil98}. 
The capture rates $W$ are calculated according to 
        \begin{equation}\label{W} 
        W=4P(\alpha Zm'_\mu)^3\frac{2J_f+1}{2J_i+1}\left(1-\frac{Q}{m_\mu+ 
        AM}\right)Q^2, 
        \end{equation} 
where $\alpha$ is the fine-structure constant, $m'_\mu$ is the reduced muon 
mass, and $Q$ is the $Q$-value of the nuclear transition. The reduced 
nuclear matrix 
elements are included in $P$ (see Ref.\ \cite{mor60} for further details). 
Instead of renormalizing the axial vector coupling constant, the 
corrections are included in the effective operators. 
Therefore, the calculations are performed using the bare value $C_{\rm A}/ 
C_{\rm V}=-1.251$. 
 
 From Fig.\ \ref{rates} it can be seen that the renormalization increases the 
capture rate for all interactions, pushing it 
closer to the experimental value for both the USD, CD-Bonn and Nijm-I 
interactions, when $C_{\rm P}/C_{\rm A}$ is close to the PCAC value. 
The USD 
calculation with the bare operators yield an interval $-4.9\le C_{\rm P}/ 
C_{\rm A}\le -3.7$, far from a reasonable expectation for the value of 
this ratio. 
The ratio $C_{\rm P}/C_{\rm A}$ calculated with the renormalized CD-Bonn and 
Nijm-I one-body operators agrees almost exactly 
with the PCAC prediction. For the PCAC prediction $C_{\rm P}/C_{\rm A}\approx7$, 
the renormalized USD calculation yields a capture rate slightly below the 
experimental 
window. However, the role of the renormalization is similar to that seen with the  
Nijm-I and CD-Bonn interactions.   
As soon as the angular-correlation data on the muon 
capture in $^{20}$Ne are published, the predictions of Fig.\ \ref{xfig} can 
be used for the extraction of the ratio $C_{\rm P}/C_{\rm A}$. At this point 
we can observe that the general trend is very similar to the $^{28}$Si case 
of Ref.\ \cite{sii99}. The renormalized calculations reduce the magnitude of 
$x$ for a given $C_{\rm P}/C_{\rm A}$ ratio, and the 
behaviour is very similar for the USD, CD-Bonn \cite{mac96} and Nijm-I \cite{nim94} 
interactions. This supports 
the conclusion of Ref.\ \cite{sii99}, where the qualitative effects of the 
renormalization on the $x$ were found to be interaction independent. 
 
In conclusion, our calculations support the near interaction indepedence of the 
effects of the 
renormalization of the one-body transition operators involved in the shell-model 
calculation of the muon-capture rates and the angular-correlation parameter $x$. 
This renormalization is introduced by replacing the bare transition 
operators, operating in the full Hilbert space, by effective ones, calculated 
with the CD-Bonn and Nijm-I interactions and now  
operating in the shell-model valence space. 
In the present work we found that these effective operators give very 
satisfactory results when compared to the experimental data. This is 
confirmed by the capture rates, where the agreement with experiment is much 
better with the effective operators. We have also given predictions for the ratio 
$x=M_1(2)/M_1(-1)$, which 
can be used for the determination of the ratio $C_{\rm P}/C_{\rm A}$ as soon as 
the experimental anisotropy data becomes available. If $C_{\rm P}/C_{\rm A}\approx7$, 
as predicted by PCAC and as seen in the capture rate 
calculations, then $x\sim 0.35$ for all interactions employed. The results from 
$^{28}$Si indicate however \cite{bru99,sii99} that $C_{\rm P}/C_{\rm A}\sim 5$. 
The latter value would yield $x\sim 0.30$ for 
the present reaction. With  $C_{\rm P}/C_{\rm A}\sim 5$, 
the capture rates reported in Fig.\ \ref{rates} will not deviate 
much from experiment. An experimental determination of $x$ would then 
clarify this situation.

\clearpage

\begin{figure} 
\caption{Calculated and experimental \cite{CD} energy spectra of $^{20}$F.} 
        \label{fig:spec} 
\end{figure}

\begin{figure} 
        \caption{Capture rates leading to the $1^+_1$ (1.057 MeV) final state 
        in $^{20}$F.} 
        \label{rates} 
\end{figure} 
 
\begin{figure} 
\caption{Parameter $x$ of Eq.\ (\ref{x}) 
                 plotted as a function of the ratio $C_{\rm P}/C_{\rm A}$.} 
        \label{xfig} 
\end{figure}

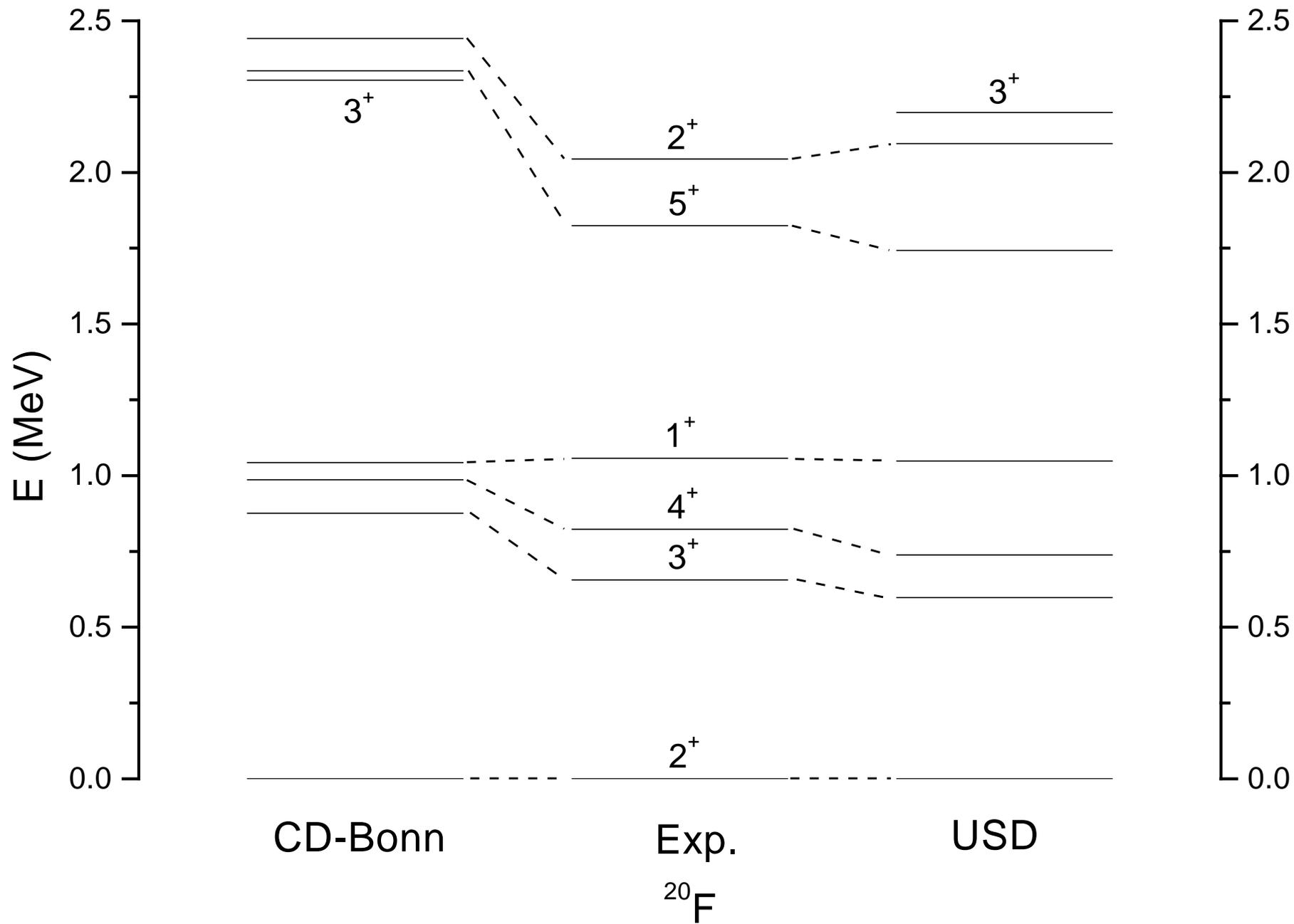
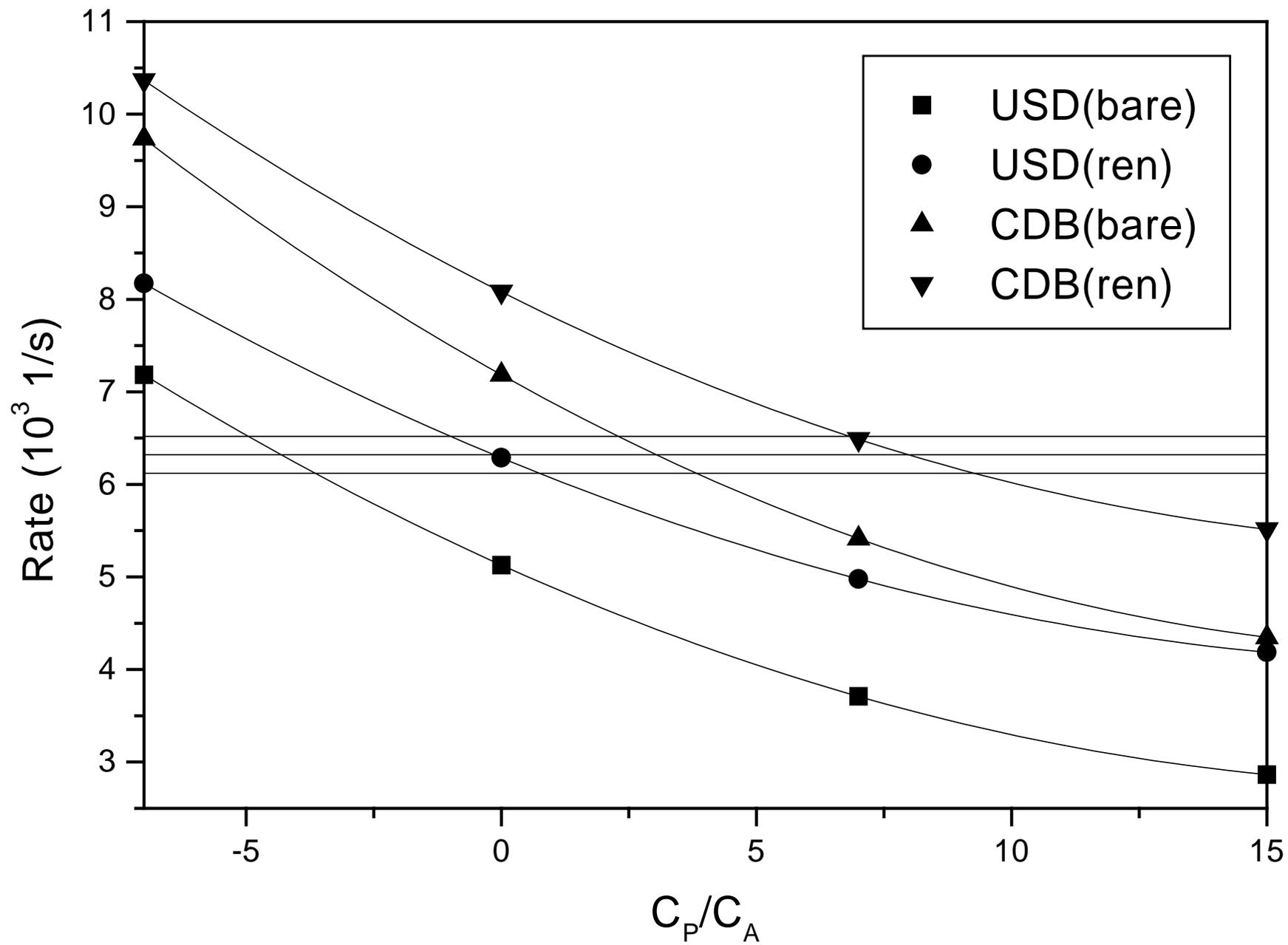
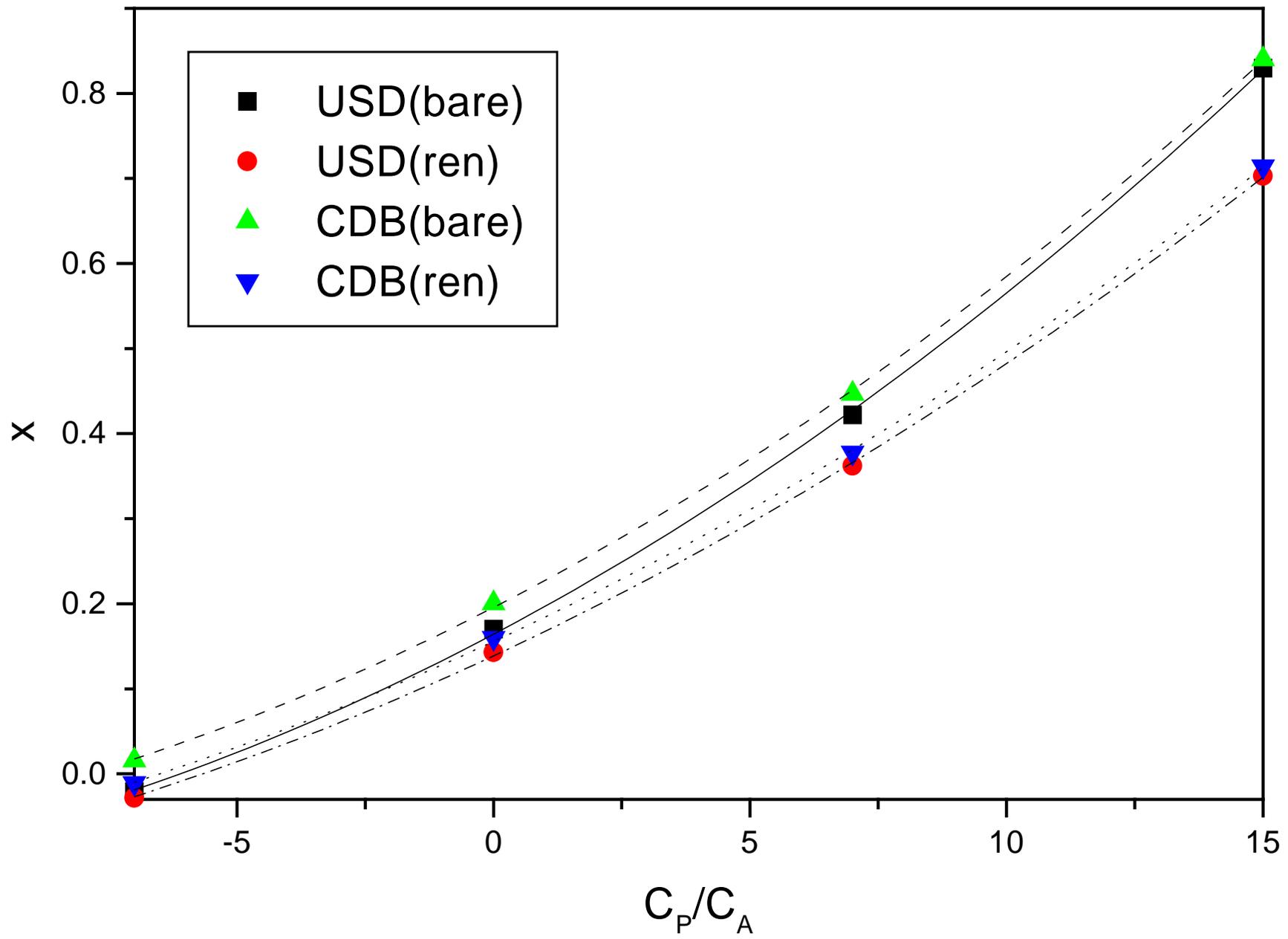
\end{document}